\documentclass[letterpaper,11pt,fleqn]{article}

\usepackage[utf8]{inputenc} 
\usepackage{fullpage,physics,graphicx,xcolor,bm,bbm,authblk, amsmath,amsfonts,amsbsy,amssymb, hyperref}
\usepackage[numbers, square, sort&compress]{natbib}
\hypersetup{colorlinks = true, linkbordercolor = white, linkcolor = blue, citecolor = blue}
\newcommand{\equa}[1]{Eq.~(\ref{#1})} \newcommand{\equas}[1]{Eqs.~(\ref{#1})}
\newcommand{\equass}[2]{Eqs.~(\ref{#1})-(\ref{#2})}
\newcommand{\equasa}[2]{Eqs.~(\ref{#1}) and (\ref{#2})}
\newcommand{\Pe}{{\rm Pe}} \newcommand{\Rey}{{\rm Re}} \newcommand{\Ge}{{\rm Ge}} 
\newcommand{\eqn}[2]{\begin{gather}
#1
\label{#2}
\end{gather}
}
\newcommand{\gat}[2]{\begin{subequations}\label{#2}\begin{gather}
#1
\end{gather}\end{subequations}
}
\newcommand{\pg}{\hphantom{0}}

\title{\bf Dissipation instability of Couette-like adiabatic flows in a plane channel}
\author{\bf A. Barletta\footnote{Corresponding author: \texttt{antonio.barletta@unibo.it}}\ $^{, 1}$; M. Celli$^1$; S. Lazzari$^2$; P. V. Brandão$^1$}
\affil{\small $^1$Department of Industrial Engineering, Alma Mater Studiorum Universit\`a di Bologna,\\
Viale Risorgimento 2, 40136 Bologna, Italy\\
\vspace{2mm}
$^2$Department of Architecture and Design, University of Genoa,\\ Stradone S. Agostino 37, 16123 Genoa, Italy
}

\date{\small\today} 

\begin{document}

\maketitle

\begin{abstract}
\noindent 
The mixed convection flow in a plane channel with adiabatic boundaries is examined. The boundaries have an externally prescribed relative velocity defining a Couette-like setup for the flow. 
A stationary flow regime is maintained with a constant velocity difference between the boundaries, considered as thermally insulated. The effect of viscous dissipation induces a heat source in the flow domain and, hence, a temperature gradient. The nonuniform temperature distribution causes, in turn, a buoyancy force and a combined forced and free flow regime.
Dual mixed convection flows occur for a given velocity difference. Their structure is analysed where, in general, only one branch of the dual flows is compatible with the Oberbeck-Boussinesq approximation, for realistic values of the Gebhart number. A linear stability analysis of the basic stationary flows with viscous dissipation is carried out. The stability eigenvalue problem is solved numerically, leading to the determination of the neutral stability curves and the critical values of the P\'eclet number, for different Gebhart numbers.  An analytical asymptotic solution in the special case of perturbations with infinite wavelength is also developed.
\\[0.5cm]
\end{abstract}

\newpage

\section{Introduction}
The absence of a transition to hydrodynamic instability, through linear perturbations,  for the plane Couette flow is a cornerstone result of fluid mechanics. In fact, early discussions regarding the linear stability of the plane Couette flow date back to papers such as that by \citet{rayleigh1914lxii}, while a rigorous proof was provided in more recent times by \citet{romanov1973stability}. The core assumption of this important achievement is that the fluid is considered isothermal so that no temperature gradient effect may influence the local momentum balance of the fluid. A thorough discussion of the hydrodynamic stability for the plane Couette flow is provided, for instance, in the books by \citet{drazin2004hydrodynamic} and by \citet{schmid2000stability}. Despite the theoretical results of the linear stability analysis, transition to instability and turbulence as a response to finite-amplitude perturbations is observed in several experiments such as those reported by \citet{bottin1998experimental} and by \citet{tillmark1992experiments}. Transition to instability was indeed proved experimentally for Reynolds numbers around 300.

When a non-isothermal regime is considered, convection heat transfer may cause the linear instability of the plane Couette flow at sufficiently large Rayleigh numbers, where the Rayleigh number is proportional to the temperature difference imposed between the hot lower wall and the cold upper wall \cite{ingersoll1966convective, kimura1971convective, KELLY199435}. The thermal buoyancy force is the cause of the motion by means of an upward heat flux imposed through the temperature boundary conditions. Besides the boundary conditions, a temperature gradient inside the fluid may be caused by the frictional heating. In fact, a viscous dissipation effect occurs due to the velocity difference between the plane boundaries.

The effect of viscous dissipation may be the unique source of thermal instability in those cases where there is no temperature difference impressed on the fluid by means of the boundary conditions. In fact, the temperature coupling term within the local momentum balance equation can be manifold. The most common coupling terms are the viscous force, since the viscosity depends significantly on the local temperature, and the buoyancy force, since the density depends significantly on the local temperature. The variable viscosity, in connection with viscous dissipation, has been envisaged as a cause of thermally-induced flow instability by \citet{joseph1964variable, joseph1965stability} and, more recently, by \citet{barletta2012variable}. The buoyancy force, as responsible of a viscous dissipation instability, has been also considered in several studies over the last decades \cite{white2002experimental, barletta2010convection, barletta2011onset, barletta2015thermal, requile2020viscous, doi0144878}. Obviously, the physics of the viscous dissipation effect suggests that both the temperature-dependence of the fluid viscosity and the temperature-dependence of the fluid density may contribute in some way to the momentum balance with a relative importance that may vary from case to case. 
Such a view is virtually compatible with the Oberbeck-Boussinesq model for buoyant flows, as this approximate scheme may include also cases where the viscosity or other fluid properties, such as the thermal diffusivity, undergoes temperature changes \cite{capone1994nonlinear, capone1995nonlinear, fusi2023rayleigh}. The problem is that such an expanded version of the Oberbeck-Boussinesq model is highly complicated by a significantly large number of governing parameters. Recently, some early attempts to define the methods for the investigation of flows with a large number of governing parameters have been made, based on machine learning techniques \cite{singh2023predicting}. We are guided by the principle that understanding the basic nature of physical phenomena needs a clear view of appropriate, though simplified, ontologies \cite{
horsch2020reliable}. 
For such reasons, the focus of the analysis presented in this paper will be on the classical Oberbeck-Boussinesq approximation. The idea of modelling the fluid viscosity as constant is appropriate as the focus of the Oberbeck-Boussinesq approximation is on convection processes where small temperature changes arise \cite{rajagopal1996oberbeck, BARLETTA2022103939, barletta2023use}.  

The aim of this paper is bringing a different view on the onset of instability for the Couette flow. Indeed, the term Couette-like flow is more appropriate as we will show that taking into account the effects of viscous dissipation and of the buoyancy force yields a modification of the Couette profile in the stationary basic flow conditions. The arrangement of the Couette boundary conditions specifies adiabatic walls subjected to a velocity difference, where adiabaticity ensures the absence of an external thermal forcing. Non-isothermal flow occurs as a consequence of viscous friction which is, in turn, a consequence of the impressed velocity difference between the boundary walls. This paper is intended as an extension to a Couette-like flow system of the analysis carried out for a Poiseuille-like flow system in a recent study \cite{doi0144878}. The analysis of the transition to instability of the basic flow with linear perturbations is carried out under conditions where the viscous dissipation effect is likely to yield major effects, namely the creeping flow of a fluid with a very large Prandtl number \cite{barletta2010convection, doi0144878}. 

\section{Viscous Dissipation Buoyant Flow}

\begin{figure}[t]
\centering
\includegraphics[width=0.75\textwidth]{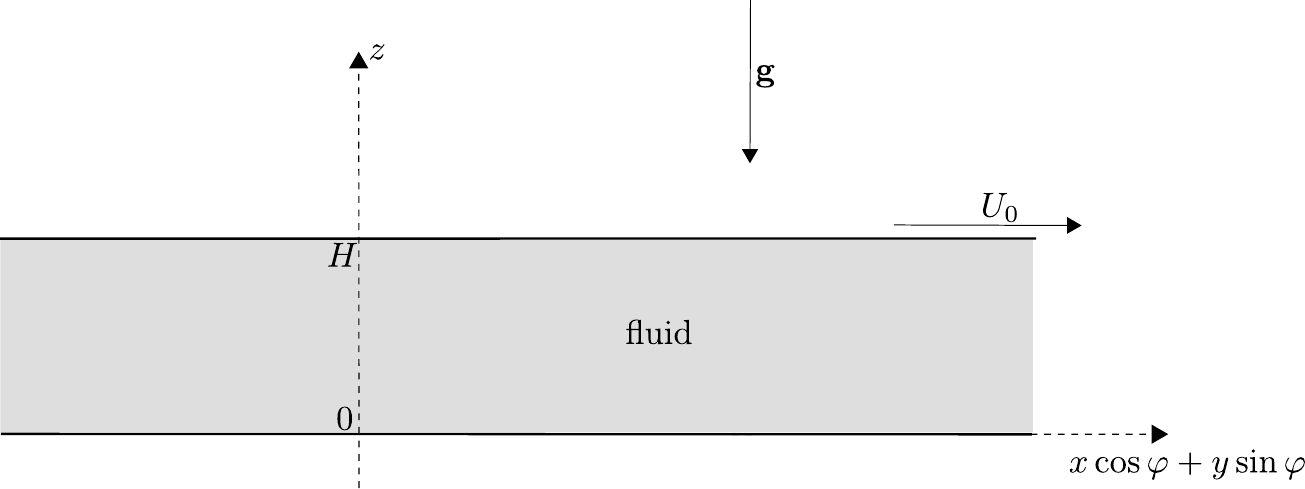}
\caption{\label{fig1}A sketch of the flow system}
\end{figure}

Following the classical Couette-like setup, we consider the Newtonian flow within a plane-parallel channel caused by the relative velocity $U_0$ between the boundary walls at $z = 0$ and $z = H$. Here, $z$ is the vertical axis perpendicular to the walls and $H$ is the distance between the channel boundaries (see Fig.~\ref{fig1}). The width in both the horizontal $x$ and $y$ directions is assumed as infinite. The uniform gravitational acceleration $\vb{g}$ is parallel to the $z$ axis, so that $\vb{g} = - g\, \vu{e}_z$, where $\vu{e}_z$ is the unit vector of the $z$ axis and $g$ is the modulus of $\vb{g}$. 

The boundary conditions prescribed at $z=0, H$ correspond to impermeable and perfectly adiabatic walls with an imposed relative velocity,
\eqn{
\vb{u} = 0 \qc \pdv{T}{z} = 0 \qfor z=0,
\nonumber\\
u = U_0 \, \cos\varphi \qc v = U_0\, \sin\varphi \qc w = 0 \qc \pdv{T}{z} = 0 \qfor z=H,
}{3}
where $\vb{u} = (u, v, w)$ is the velocity field and $T$ is the temperature field, while $U_0$ is the velocity of the upper boundary wall in the direction defined by the unit vector $\qty(\cos\varphi, \sin\varphi, 0)$, with $0 \le \varphi \le \pi/2$.

\subsection{Governing Equations}
Within the framework of the Oberbeck-Boussinesq approximation, the governing equations are given by
\gat{
\div{\vb{u}} = 0, \label{1a}\\
\pdv{\vb{u}}{t} + (\vb{u} \cdot \grad{)\,\vb{u}} = - \frac{1}{\rho}\, \grad{p} + g \beta \qty(T - T_0)\, \vu{e}_z + \nu\, \nabla^2 \vb{u},\label{1b}\\
\pdv{T}{t} + (\vb{u} \cdot \grad{)\,T} = \alpha \, \nabla^2 T + \frac{\nu}{c}\, \Phi , \label{1c}
}{1}
where $\rho$, $\beta$, $\nu$, $\alpha$ and $c$ are the fluid density, thermal expansion coefficient, kinematic viscosity, thermal diffusivity and specific heat of the fluid in the reference thermodynamic state with constant temperature $T_0$. In \equas{1}, $t$ is the time and $p$ is the local difference between the pressure and the hydrostatic pressure. 
The dissipation function $\Phi$, employed in \equa{1c} denotes 
\eqn{
\Phi = \frac{1}{2} \, \gamma_{ij} \gamma_{ij} \qusing \gamma_{ij} = \pdv{u_i}{x_j} + \pdv{u_j}{x_i} ,
}{2}
where Einstein's notation for the implicit sum over repeated indices is used, $\gamma_{ij}$ is the $ij$ Cartesian component of shear rate tensor, while $u_i$ and $x_i$ denote the $i$th components of the velocity vector $\vb{u}$ and of the position vector $\vb{x} = (x,y,z)$, respectively.

\subsection{Dimensionless Formulation}
A dimensionless formulation of \equasa{3}{1} can be achieved by the scaling
\eqn{
\frac{\vb{x}}{H} = \frac{(x,y,z)}{H} \to (x,y,z) = \vb{x} \qc \frac{t}{H^2/\alpha} \to t \qc \frac{\vb{u}}{\alpha/H} =\frac{(u,v,w)}{\alpha/H} \to (u,v,w) = \vb{u} , \nonumber\\ 
\frac{p}{\rho \alpha \nu/H^2} \to p \qc \frac{T - T_0}{\Delta T} \to T \qc \frac{\Phi}{\alpha^2/H^4} \to \Phi \qusing \Delta T = \frac{\alpha \nu}{g \beta H^3} .
}{4}
By employing \equa{4}, one can rewrite \equas{1} in a dimensionless form,
\gat{
\div{\vb{u}} = 0, \label{5a}\\
\frac{1}{\Pr} \left[ \pdv{\vb{u}}{t} + (\vb{u} \cdot \grad{)\,\vb{u}} \right] = - \grad{p} + T\, \vu{e}_z + \nabla^2 \vb{u},\label{5b}\\
\pdv{T}{t} + (\vb{u} \cdot \grad{)\,T} = \nabla^2 T + \Ge\, \Phi , \label{5c}
}{5}
while the dimensionless boundary conditions (\ref{3}) are given by
\eqn{
\vb{u} = 0 \qc \pdv{T}{z} = 0 \qfor z=0,
\nonumber\\
u = \Pe \, \cos\varphi \qc v = \Pe\, \sin\varphi \qc w = 0 \qc \pdv{T}{z} = 0 \qfor z=1.
}{6}
In \equas{5b}, (\ref{5c}) and (\ref{6}), the Prandtl number, $\Pr$, the Gebhart number, $\Ge$, and the P\'eclet number, $\Pe$, are defined as
\eqn{
\Pr = \frac{\nu}{\alpha} \qc \Ge = \frac{g \beta H}{c} \qc \Pe = \frac{U_0 H}{\alpha}  .
}{7}

\begin{figure}[t]
\centering
\includegraphics[width=0.82\textwidth]{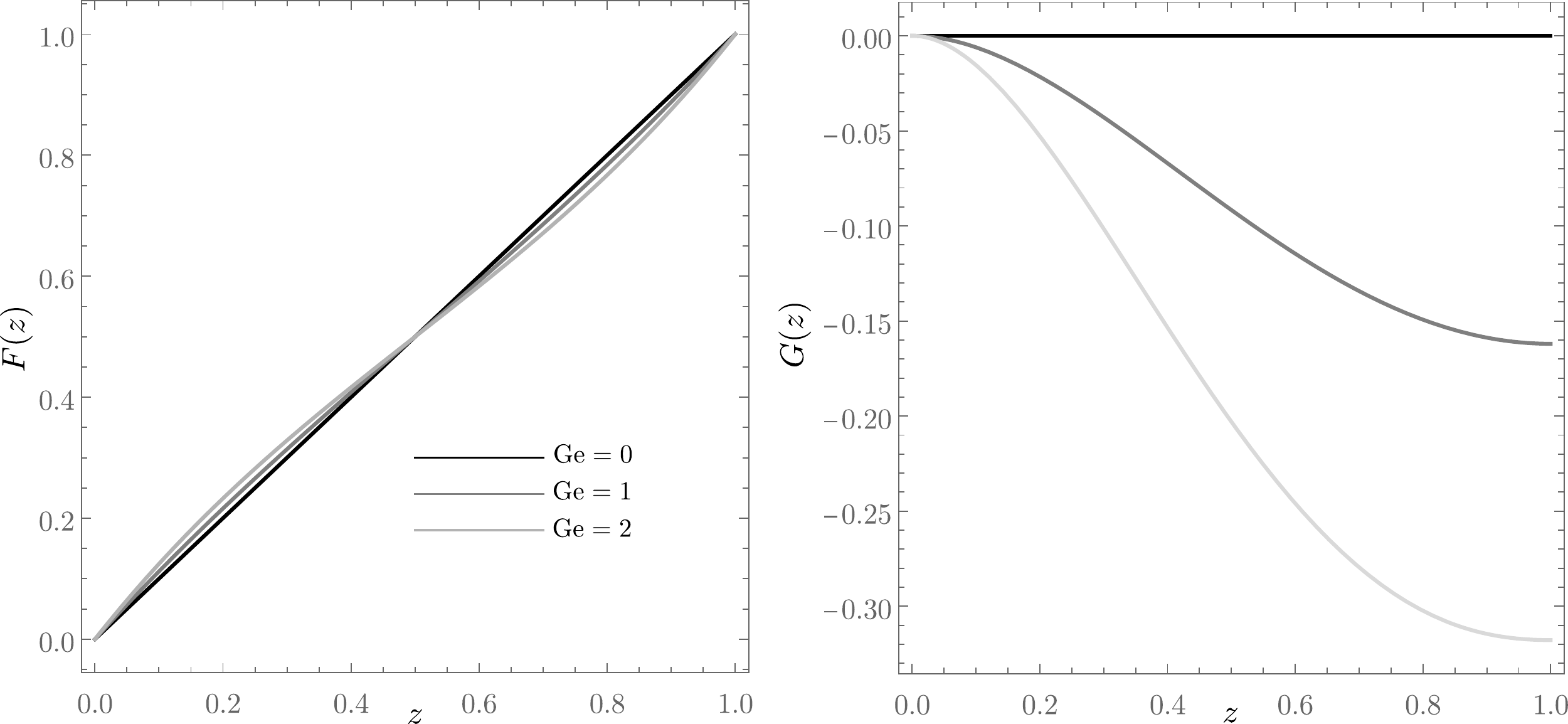}
\caption{\label{fig2}Plots of $F(z)$ and $G(z)$ for the $A = A_{-}$ branch with different values of $\Ge$}
\end{figure}

\begin{figure}[t]
\centering
\includegraphics[width=0.82\textwidth]{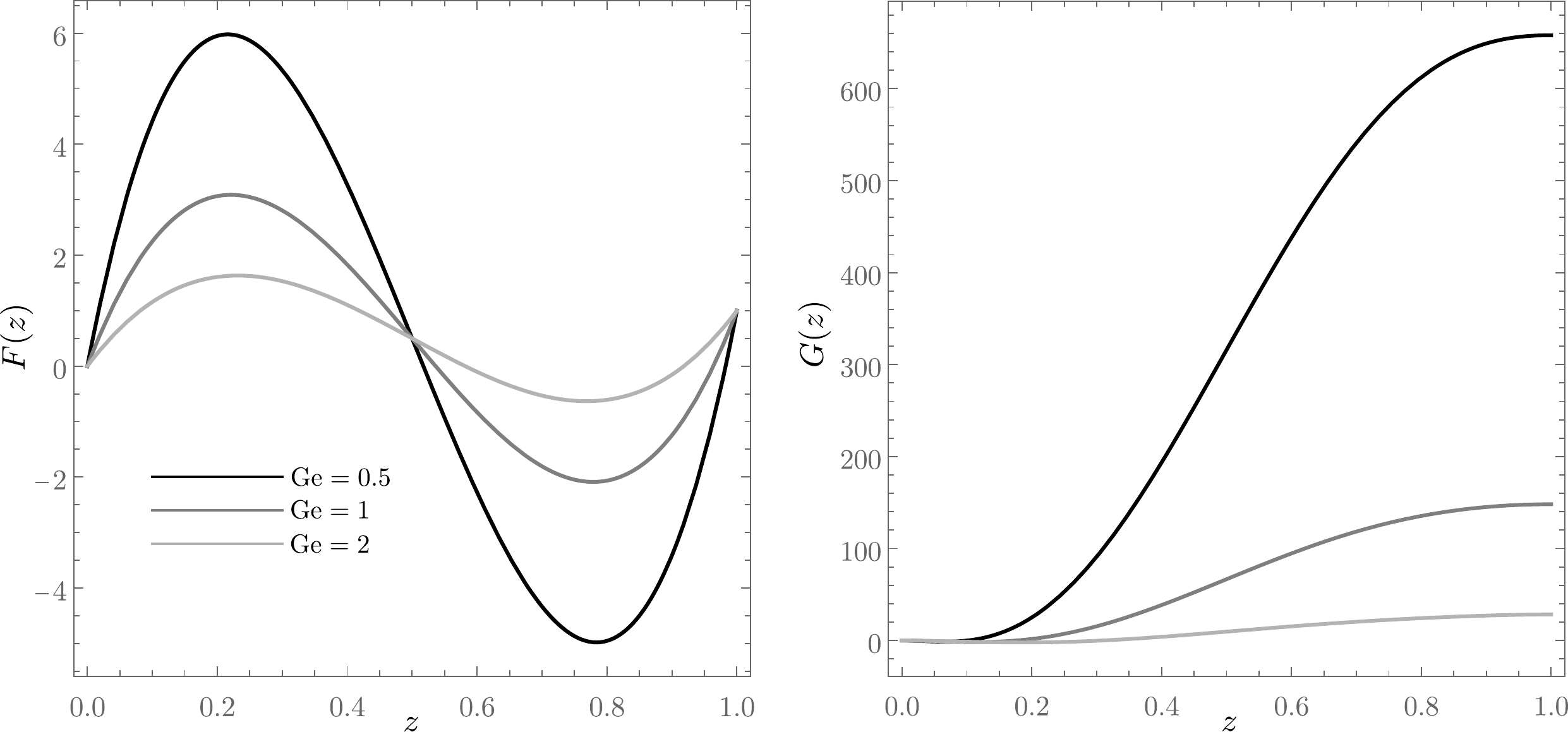}
\caption{\label{fig3}Plots of $F(z)$ and $G(z)$ for the $A = A_{+}$ branch with different values of $\Ge$}
\end{figure}

\section{Dual Adiabatic Flows}\label{badufl}
Steady-state flows satisfying \equasa{5}{6} do exist with a velocity field parallel to the unit vector $\qty(\cos\varphi, \sin\varphi, 0)$, namely
\eqn{
u_b = \Pe\, F(z)\, \cos\varphi \qc v_b = \Pe\, F(z)\, \sin\varphi \qc w_b = 0, \nonumber\\
T_b = \Pe\, A \left( x \, \cos\varphi + y\, \sin\varphi \right) + \Pe^2\, G(z), \nonumber\\
\grad{p_b} = \Big( \Pe\, F''(z)\, \cos\varphi ,\ \Pe\, F''(z)\, \sin\varphi ,\
T_b \Big) .
}{8}
Here, $b$ is a subscript meant to indicate the ``basic solution'' and primes are used for the derivatives with respect to $z$. Equations (\ref{5}) and (\ref{6}) are satisfied provided that functions $F(z)$ and $G(z)$ are the polynomials 
\eqn{
F(z) = z + \frac{A}{12} \, z \, \qty(z - 1) \qty(2 z - 1), \nonumber\\
G(z) = \frac{z^2}{1440}\, \big\{ 2 A z \left[A \left(6 z^2-15 z+10\right)+120\right] - \Ge \big[A^2 \left(12 z^4-36 z^3+40 z^2-20 z+5\right)
\nonumber\\
\hspace{2.5cm}+120 A (z-1)^2+720\big] \big\},
}{9}
while the constant $A$ is equal either to $A_{-}$ or to $A_{+}$, defined as
\eqn{
A_{-} = 12 \, \frac{ 15 - \sqrt{5 \left( 45-\Ge^2 \right)}}{\Ge} \qc
A_{+} = 12 \, \frac{ 15 + \sqrt{5 \left( 45-\Ge^2 \right)}}{\Ge} .
}{10}
It is easily verified that $F'''(z) = A$. This result implies that the basic temperature gradient in the horizontal $x$ and $y$ directions is independent of $z$.
Equations~(\ref{8})-(\ref{10}) describe horizontal flows 
where the velocity field is inclined an angle $\varphi$ to the $x$ axis. In fact, \equa{9} yields 
\eqn{
\int\limits_0^1 F(z) \, \dd z = \frac{1}{2} ,
}{11}
which means that the average dimensionless velocity in the flow direction is equal to $\Pe/2$, which is the mean velocity of the boundary walls. In other words, in the reference frame where the mean velocity of the boundary walls is zero, the average velocity of the fluid is zero and, hence, also the flow rate is zero. In the absence of viscous dissipation, such a constraint is characteristic of the Couette flow.

On account of \equa{10}, we may have $A = A_{-}$ or $A=A_{+}$. As a consequence, \equasa{8}{9} entail the existence of dual flows corresponding to the same prescribed values of $\Pe$ and $\Ge$. These dual flows are allowed only with $\Ge \le 3 \sqrt{5} \approx 6.70820$. With $\Ge = 3 \sqrt{5}$, $A_{-}=A_{+}$ and the dual flows coincide. It must be mentioned that this maximum Gebhart number is an extremely large value for any real-world system. 

\begin{figure}[t]
\centering
\includegraphics[width=0.82\textwidth]{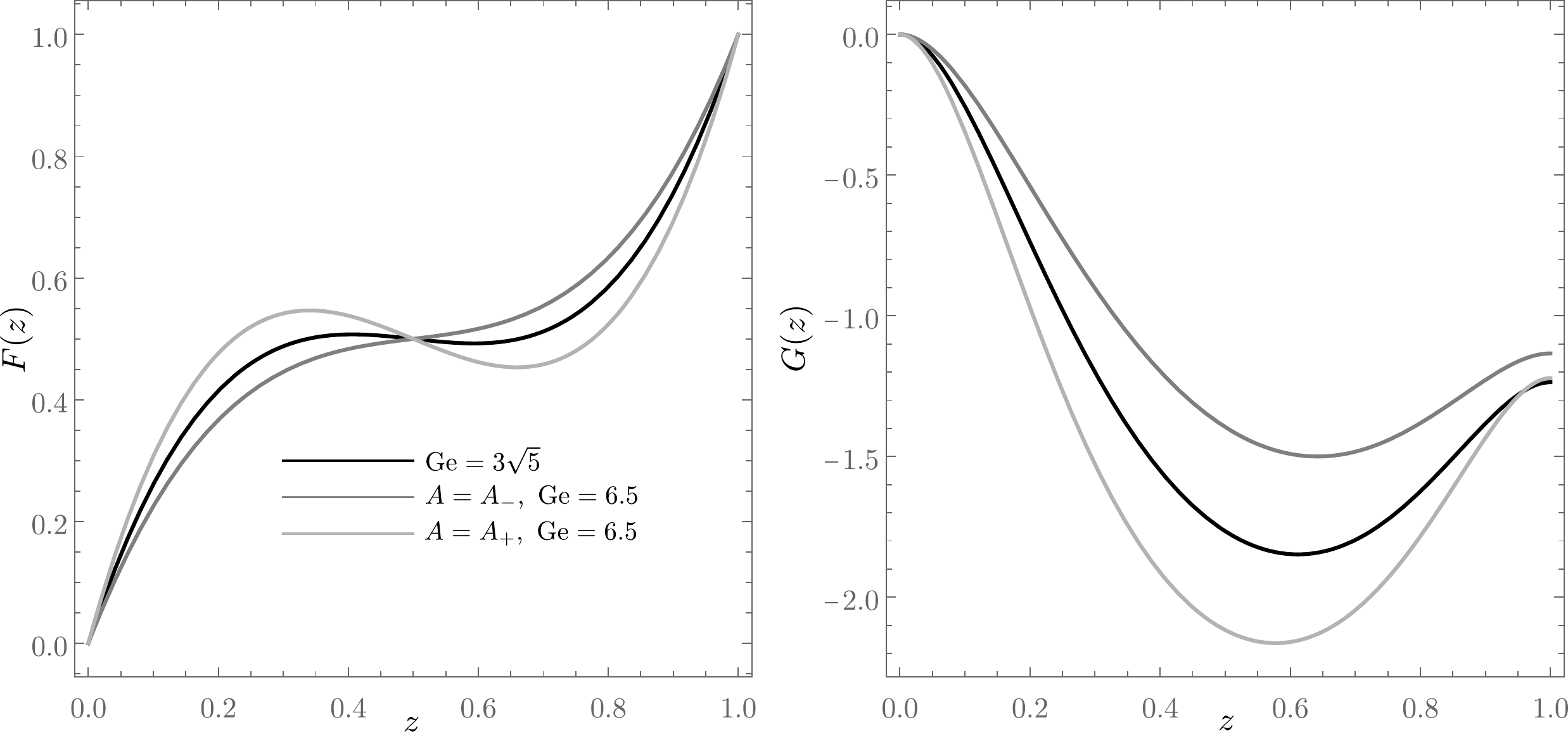}
\caption{\label{fig4}Plots of $F(z)$ and $G(z)$ for the $A = A_{-}$ branch and the $A = A_{+}$ branch with an extremely large value of $\Ge$ close to the maximum, $\Ge = 3\sqrt{5}$}
\end{figure}

By employing \equa{8}, one can infer that function $F(z)$ yields the basic velocity profile with the P\'eclet number being an overall constant factor. Similarly, with the feature $G(0)=0$ displayed by \equa{9}, the function $G(z)$ yields (up to an overall factor $\Pe^2$) the temperature difference between a given position $z$ and the bottom boundary, $z=0$, in the basic state for fixed $x$ and $y$. With these considerations in mind, one may view the plots of $F(z)$ and $G(z)$ as representations of the velocity and temperature distributions on a transverse, $x \cos\varphi + y \sin\varphi = constant$, cross-section of the channel. An interesting characteristic is that both $F(z)$ and $G(z)$ depend on a single parameter, $\Ge$. 

Figures~\ref{fig2} and \ref{fig3} show some plots of $F(z)$ and $G(z)$ with different Gebhart numbers, relative to either the $A_{-}$ branch or the $A_{+}$ branch. The flow conditions in the two solution branches are utterly different either with respect to the velocity or to the temperature profiles. Figure~\ref{fig2}, relative to the $A_{-}$ branch, displays just slight changes from the linear velocity profile of the isothermal Couette flow. Values such as $\Ge = 1$ or $2$ are very large for most applications except for geophysical or astrophysical systems. The influence of an increasing Gebhart number in the $A_{-}$ branch is more marked for the temperature profiles $G(z)$. 
Figure~\ref{fig3} shows a much more significant influence of the Gebhart number on examining the velocity and temperature profiles for the $A_{+}$ branch. Here, the similarity to the Couette linear profile is almost absent for all the considered Gebhart numbers. Such velocity profiles describe a bidirectional flow with $F(z)$ changing from positive to negative across the channel cross-section.

We mention that taking $\Ge = 0$ means switching off the effect of viscous dissipation as it comes out from \equa{5c}. This limiting case yields a linear Couette velocity profile with a uniform temperature distribution as displayed by the black lines in Fig.~\ref{fig2}. With regard to the behaviour for small Gebhart numbers, the $A_{-}$ branch shows the asymptotic expressions
\eqn{
A = A_{-} = 2\, \Ge + \order{\Ge^3} ,
\nonumber\\
F(z) = z + \frac{\Ge}{6} \, z\, (z-1) (2 z-1) + \order{\Ge^3},
\nonumber\\
G(z) = \frac{\Ge}{6}\, z^2\, (2 z-3) + \frac{\Ge^2}{180} \, z^2 \left[z \left(6 z^2-45 z+70\right) - 30\right] + \order{\Ge^3} .
}{12}
 A direct consequence of \equasa{8}{12} is that the $A_{-}$ branch solution attained with $\Ge \to 0$ yields the isothermal Couette flow,
\eqn{
u_b = \Pe\, z\, \cos\varphi \qc v_b = \Pe\, z\, \sin\varphi \qc w_b = 0 \qc
T_b = 0 \qc
\grad{p_b} = 0.
}{13}
On the contrary, the $A_{+}$ branch displays a singular behaviour in the limit $\Ge \to 0$. In fact,
\eqn{
A = A_{+} = \frac{360}{\Ge} - 2\, \Ge + \order{\Ge^3} ,
\nonumber\\
F(z) = \frac{30\,z \, (z-1) (2 z-1)}{\Ge} + z - \frac{\Ge}{6} \, z\, (z-1) (2 z-1) + \order{\Ge^3},
\nonumber\\
G(z) = \frac{180\, z^3\, \qty(6 z^2-15 z+10)}{\Ge^2}-\frac{30 \, z^2 \qty(36 z^4-108 z^3+120 z^2-62 z+15)}{\Ge}
\nonumber\\
\hspace{2cm} -\, 2 \, z^2 \left(6 z^3-20 z+15\right) + \frac{\Ge\, z^2}{6}  \left(72 z^4-216 z^3+240 z^2-122 z+27\right)
\nonumber\\
\hspace{3cm}-\, \frac{\Ge^2 z^2}{180} \left(6 z^3 - 45 z^2+70 z-30\right) + \order{\Ge^3} .
}{14}
Equation~(\ref{14}) shows that the $A_{+}$ solution branch blows up when $\Ge \to 0$. In other words, we reach the reasonable conclusion that, by switching off the effect of viscous dissipation or, equivalently, by setting $\Ge=0$, there is a unique possible solution: the isothermal Couette flow (\ref{13}). The singular behaviour of the $A_{+}$ branch entails an extremely marked influence of small values of $\Ge$ on the velocity and temperature gradients in the $z$ direction. For instance, by lowering $\Ge$ from $10^{-3}$ to $10^{-6}$, one gets an amplification of the maximum temperature difference across the range $0 \le z \le 1$ of one million times. We mention that such small values of $\Ge$ are not unlikely in laboratory experiments. This scenario is quite similar to that discussed in \citet{barletta2022mixed} with reference to Poiseuille-like adiabatic flows with viscous dissipation in an adiabatic channel. Following that discussion, we share the same conclusion: the $A_{+}$ solution branch is incompatible with the assumptions behind the Oberbeck-Boussinesq approximation except for extremely large values of $\Ge$. Therefore, exactly as in the analysis presented by  \citet{doi0144878}, our investigation of linear instability will be focussed on the $A_{-}$ branch.

For the sake of completeness, Fig.~\ref{fig4} provides an illustration of what happens when $\Ge$ is close to its maximum value allowed for the existence of the basic solution (\ref{8})-(\ref{10}), {\em i.e.} $\Ge = 3\sqrt{5}$. If the two branches, $A_{-}$  and $A_{+}$, coincide at the maximum Gebhart number, they become very similar both for the velocity profile and for the temperature profile when $\Ge$ is slightly smaller than the maximum.

\section{Onset of the Instability}
The discussion provided in Section~\ref{badufl} drove the focus of this study to the branch $A_{-}$. The features of this branch, as gathered analytically from \equass{8}{10} and graphically from Fig.~\ref{fig2}, include a Couette-like shape of the velocity profile, with small departures from the linear trend, and a wide vertical range where $\partial T_b/\partial z < 0$. The latter feature suggests a potential thermal instability of the flow. If and when such potentially unstable temperature distribution actually leads to the onset of a convective instability is the aim of the forthcoming investigation.

\subsection{Small-Amplitude Perturbations}
According to the usual modal analysis of the linear instability, we perturb the basic state with small-amplitude wavelike disturbances
\eqn{
\pmqty{\vb{u}\\p\\T} = \pmqty{\vb{u}_b\\p_b\\T_b} + \varepsilon\! \pmqty{\vb{U}(z)\\P(z)\\\Theta(z)} e^{i \,k\, x} \, e^{\lambda\, t} ,
}{15}
where $\varepsilon \ll 1$ is a small positive perturbation parameter, $k$ is the wavenumber and $\lambda$ is a complex parameter. The real part of such a complex variable, $\lambda_r$, yields the temporal growth rate while  its imaginary part is $-\omega$, where $\omega$ is the angular frequency of the wave. The physical meaning of $\lambda_r$ arises when the stable or unstable behaviour of the disturbance must be assessed. The linear instability occurs when $\lambda_r > 0$, while $\lambda_r = 0$ defines the neutral stability condition. The neutral stability is the parametric threshold condition for the onset of the instability. 

We note that \equa{15} defines plane wave disturbances travelling along the horizontal $x$ direction. There is no lack of generality in this assumption as the basic flow direction is parallel to the $x\cos\varphi + y\sin\varphi$ axis, with an arbitrary angle $\varphi$ within the range $0 \le \varphi \le \pi/2$. Hence, the $x$ direction is arbitrary, relatively to the direction of the basic flow. By varying the angle, one can span all possible oblique modes of perturbation ranging from $\varphi = 0$, for the transverse modes, to $\varphi = \pi/2$, for the longitudinal modes. The Cartesian components of $\vb{U}(z)$ are denoted as $U(z)$, $V(z)$ and $W(z)$. By employing the basic solution, \equasa{8}{9}, and by substituting \equa{15} into \equasa{5}{6}, we obtain
\gat{
W' + i k\, U = 0, \label{16a}\\
\frac{1}{\Pr} \left[ \lambda\, U + i k\, \Pe \, F(z) \,U \cos\varphi + \Pe \, F'(z) \,W \cos\varphi \right] = - i k\, P + U'' - k^2\, U,\label{16b1}\\
\frac{1}{\Pr} \left[ \lambda\, V + i k\, \Pe \, F(z) \,V \cos\varphi + \Pe \, F'(z) \,W \sin\varphi \right]  = V'' - k^2 \, V ,\label{16b2}\\
\frac{1}{\Pr} \left[ \lambda\, W + i k\, \Pe \, F(z) \,W \cos\varphi  \right] = - P' + \Theta + W'' - k^2\, W, \label{16b3}\\
\lambda\, \Theta + i k\, \Pe \, F(z) \,\Theta \cos\varphi + \Pe \, A \, \qty(U \cos\varphi + V \sin\varphi) + \Pe^2\, G'(z) \, W 
\nonumber\\
\hspace{2cm}= \Theta'' - k^2\, \Theta + 2\, \Ge\, \Pe\ F'(z) \left[  \qty(U' + i k \, W) \cos\varphi + V' \sin\varphi \right] , \label{16c}
}{16}
with the boundary conditions
\eqn{
\vb{U} = 0 \qc \Theta' = 0 \qfor z=0, 1.
}{17}

\subsection{Creeping Flow}\label{creeflo}
Equations~(\ref{16}) and (\ref{17}) yield a system of homogenous ordinary differential equations with homogeneous boundary conditions, {\em i.e.}, the stability eigenvalue problem. A reasonable approximation is the assumption that the Prandtl number of the fluid is very large so that the dynamics of the perturbations is that of a creeping buoyant flow \cite{doi0144878}. In fact, a very large Prandtl number identifies a very viscous fluid with a small thermal diffusivity. Both these features, large viscosity and small thermal diffusivity, are present when the flow internal heating due to viscous dissipation is significant.

Mathematically, one takes the limit $\Pr \to \infty$ with $\Pe \sim \order{1}$ in \equass{16b1}{16b3}. Thus, \equa{16b2} simplifies to $V'' - k^2\, V = 0$. Since \equa{17} prescribes $V=0$ at $z=0,1$, the only possible solution is $V(z)=0$ for every $z$. Owing to the relation linking $U$ and $W'$, \equa{16a}, one can rearrange \equas{16b1}, (\ref{16b3}) and (\ref{16c}) in order to attain a reformulation of the stability eigenvalue problem using only the eigenfunctions $W$ and $\Theta$, namely
\gat{
W'''' - 2\,k^2\, W'' + k^4\, W - k^2\, \Theta = 0 ,\label{18a}\\
\Theta'' - \qty[k^2 + \lambda + i k\, \Pe\, F(z) \cos\varphi ]\, \Theta + \frac{2 i\, \Ge\, \Pe\, F'(z) \cos\varphi}{k}\, \qty(W'' + k^2\, W) \nonumber\\
\hspace{2cm}
-\, \frac{i\, \Pe\, A \cos\varphi}{k}\, W' - \Pe^2\, G'(z)\, W = 0  , \label{18b}\\
W = 0 \qc W' = 0 \qc \Theta' = 0 \qfor z = 0, 1. \label{18c}
}{18}
With the aim of determining the neutral stability condition, the real part of $\lambda$ is set to zero so that $\lambda = - i \omega$. 
%

\begin{figure}[t]
\centering
\includegraphics[width=0.41\textwidth]{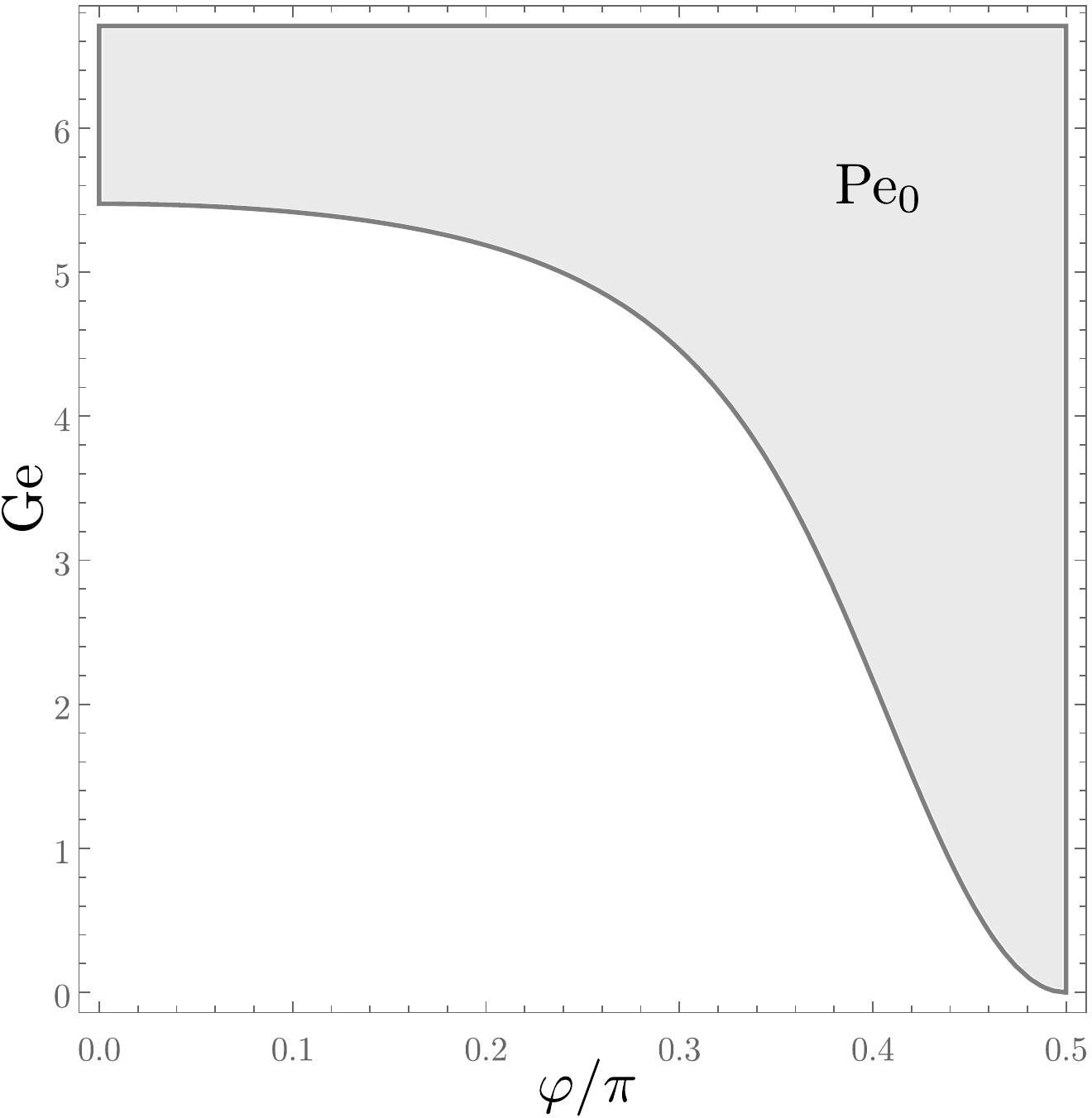}
\caption{\label{fig5}Domain of existence (grey region) of a neutral stability condition, $\Pe = \Pe_0$,  with $A = A_{-}$ and $k \to 0$}
\end{figure}

\subsection{Infinite Wavelength Perturbations}\label{infwape}
An asymptotic solution of \equas{18} can be sought for very small wavenumbers. A convenient reformulation of \equas{18} is obtained by defining 
\eqn{
\hat{\Theta} = k \, \Theta ,
}{22}
so that one may write
\gat{
W'''' - 2\,k^2\, W'' + k^4\, W - k\, \hat{\Theta} = 0 ,\label{23a}\\
\hat{\Theta}'' - \qty[k^2 - i \omega + i k\, \Pe\, F(z) \cos\varphi ]\, \hat{\Theta} + 2 i\, \Ge\, \Pe\, F'(z) \, \qty(W'' + k^2\, W) \cos\varphi \nonumber\\
\hspace{2cm} -\, i\, \Pe\, A \, W' \cos\varphi - k\, \Pe^2\, G'(z)\, W = 0  , \label{23b}\\
W = 0 \qc W' = 0 \qc \hat{\Theta}' = 0 \qfor z = 0, 1. \label{23c}
}{23}

\begin{figure}[t]
\centering
\includegraphics[width=0.41\textwidth]{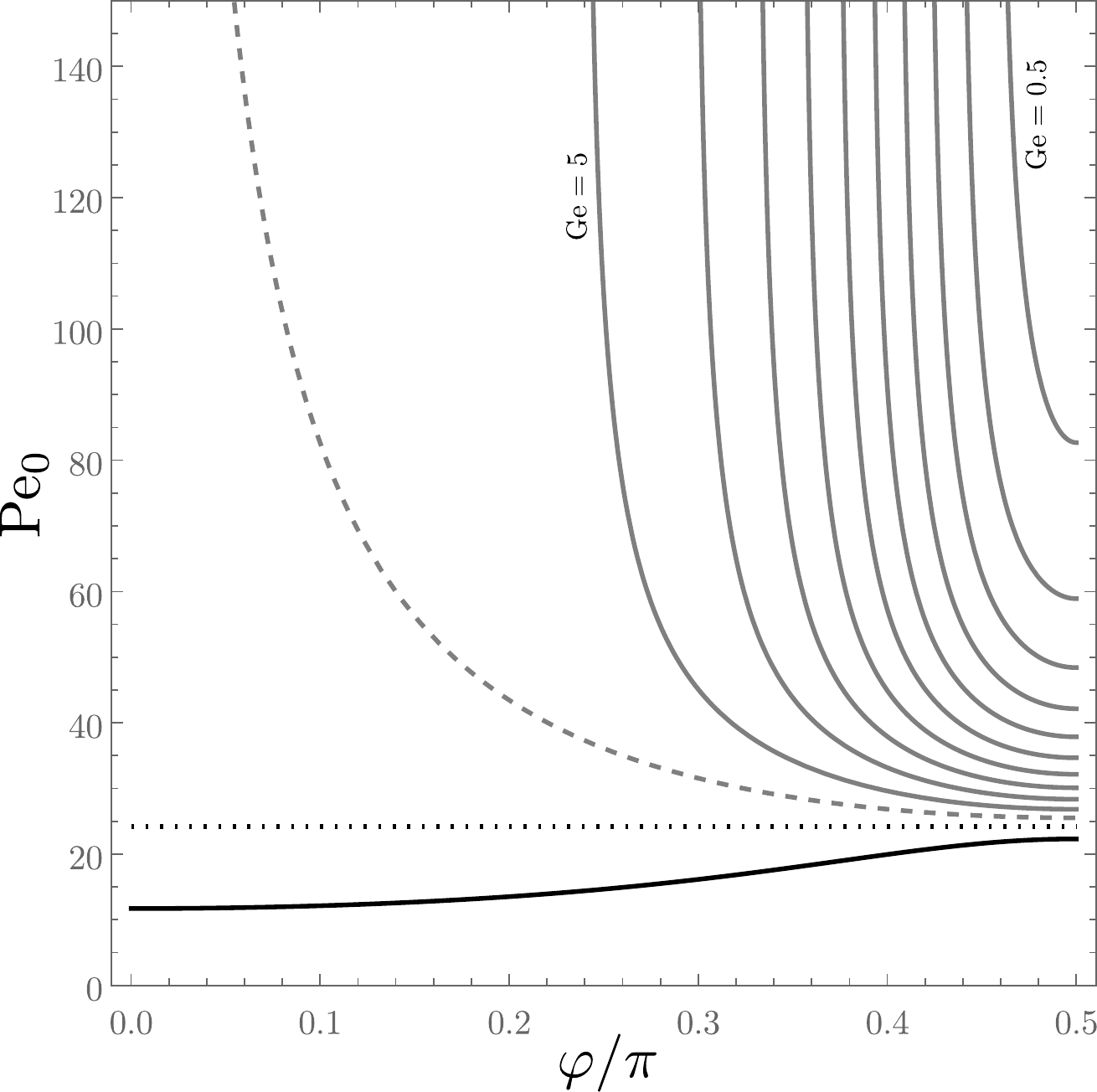}
\caption{\label{fig6}Plots of $\Pe_0$ versus $\varphi$,  with $A = A_{-}$. The solid grey lines are from $\Ge = 0.5$ to $\Ge = 5$ in steps of $0.5$. The dashed line is for $\Ge = 5.47464$, the dotted line is for $\Ge = 5.99306$ and the solid black line is for $\Ge = 3 \sqrt{5}$}
\end{figure}

\noindent{}We now consider the series expansions
\eqn{
W(z) = W_0(z) + W_1(z) \, k + W_2(z) \, k^2 + \order{k^3},
\nonumber\\
\hat{\Theta}(z) = \hat{\Theta}_0(z) + \hat{\Theta}_1(z) \, k + \hat{\Theta}_2(z) \, k^2 + \order{k^3},
\nonumber\\
\Pe = \Pe_0 + \Pe_1 \, k + \Pe_2 \, k^2 + \order{k^3},
\nonumber\\
\omega =  \omega _0 + \omega_1 \, k + \omega_2 \, k^2 + \order{k^3},
}{24}
we substitute them into \equas{23} and separately solve the differential problems to every order $k^n$ with $n=0,1,2, \ldots\ $. To order $k^0$, we just obtain
\eqn{
W_0(z) = 0 \qc \hat{\Theta}_0(z) = 1 \qc \omega_0 = 0.
}{25}
We note that $ \hat{\Theta}_0(z)$ could have been set equal to any real or complex constant. In fact, choosing $\hat{\Theta}_0(z) = 1$ is just the simplest way to break the scale invariance for the solution of a homogeneous boundary value problem by imposing the constraint $\hat{\Theta}(0) = 1$. Hence, for every $n > 0$, one has the extra boundary condition $\hat{\Theta}_n(0) = 0$. On account of \equa{25}, the boundary value problem to order $k^1$ can be written as
\gat{
W_1'''' - 1 = 0 ,\label{26a}\\
\hat{\Theta}_1'' + i\, \qty[ \omega_1 - \Pe_0\, F(z) \cos\varphi ] + 2 i\, \Ge\, \Pe_0\, F'(z) \, W_1''  \cos\varphi 
- i\, \Pe_0\, A \, W_1' \cos\varphi = 0  , \label{26b}\\
W_1(0) = 0 \qc W_1'(0) = 0 \qc \hat{\Theta}_1(0) = 0 \qc \hat{\Theta}_1'(0) = 0 , \nonumber\\
W_1(1) = 0 \qc W_1'(1) = 0 .  \label{26c}
}{26}
The solution can be easily found though we omit here the expressions of $W_1(z)$ and $\hat{\Theta}_1(z)$ for the sake of brevity. We just mention that the boundary condition $\hat{\Theta}_1'(1) = 0$ is not needed to determine the unique solution of \equas{26}, but imposing such an extra condition allows one to  write 
\eqn{
\omega_1 = \frac{\Pe_0\, \qty( 180 - A \,\Ge) \cos\varphi }{360}  ,
}{27}
with $\Pe_0$ yet undetermined.
On account of \equa{25}, the boundary value problem to order $k^2$ is given by
\gat{
W_2'''' - \hat{\Theta}_1 = 0 ,\label{28a}\\
\hat{\Theta}_2'' - 1 + i \omega_2 - i \, \Pe_1\, F(z) \cos\varphi 
+ i\, \qty[\omega_1 - \Pe_0\, F(z) \cos\varphi ]\, \hat{\Theta}_1 
\nonumber\\
\hspace{2cm} 
+\, 2 i\, \Ge\, \Pe_1\, F'(z) \, W_1''  \cos\varphi + 2 i\, \Ge\, \Pe_0\, F'(z) \, W_2''  \cos\varphi 
\nonumber\\
\hspace{3cm} 
-\, i\, \Pe_1\, A \, W_1' \cos\varphi - i\, \Pe_0\, A \, W_2' \cos\varphi - \Pe_0^2\, G'(z)\, W_1 = 0  , \label{28b}\\
W_2(0) = 0 \qc W_2'(0) = 0 \qc \hat{\Theta}_2(0) = 0 \qc \hat{\Theta}_2'(0) = 0 , \nonumber\\
W_2(1) = 0 \qc W_2'(1) = 0 .  \label{28c}
}{28}
Again, we omit the explicit expressions of $W_2(z)$ and $\hat{\Theta}_2(z)$ for the unique solution of \equas{28}. We only stress that the extra boundary condition,  $\hat{\Theta}_2'(1) = 0$, not involved in the boundary value problem~(\ref{28}) yields an explicit and unique expression for a positive $\Pe_0$, 
\eqn{
\Pe_0 = 720\, \sqrt{154} \; \Big\{2 \, A \, (A+66)\, \Ge^2 \cos (2 \varphi ) - 220\, \qty[(A-81) \,A + 1512] \cos (2 \varphi )
\nonumber\\
\hspace{2cm}
+\, A \, \qty[2\, (A+66) \,\Ge^2 + 77\, A\, \Ge +220\, (9 - 2\, A) ] + 55440\, (\Ge - 6) \Big\}^{-1/2} .
}{29}
It must be stressed that $\Pe_0$ has a very important physical meaning. As a matter of fact, $\Pe_0$ yields the neutral threshold for linear instability with perturbation normal modes having $k \to 0$.

Equation~(\ref{29}), which holds both for $A = A_{-}$ and $A = A_{+}$, gives a real positive value of $\Pe_0$ only for values of $\Ge$ and $\varphi$ such that the expression in curly brackets on the right hand side is non-negative. However, this condition does not hold for every possible pair $\qty(\Ge, \varphi)$. 

Figure~\ref{fig5} shows the domains of existence for $\Pe_0$, evaluated through \equa{29}, by considering either the branch $A = A_{-}$ or the branch $A = A_{+}$. An interesting fact is that, by setting $A = A_{-}$, $\Pe_0$ exists for longitudinal modes $(\varphi = \pi/2)$ within the whole range $0 < \Ge \le 3\sqrt{5}$. For every other value of $\varphi$, there is always a minimum $\Ge$ below which $\Pe_0$ does not exist. Some plots of $\Pe_0$ versus $\varphi$, also relative to the choice $A = A_{-}$, are reported in Fig.~\ref{fig6} for different values of $\Ge$. In this figure, there are grey lines relative to $\Ge$ from $0.5$ to $5$ in steps of $0.5$, while the line of maximum $\Ge$ is drawn as a black line. There are also a dotted line and a dashed line corresponding to a couple of special values of $\Ge$. The dotted line, for $\Ge = 5.99306$, identifies a special case where $\Pe_0$ is independent of $\varphi$ and has the value $24.2005$. The dashed line is relative to $\Ge = 5.47464$, which is the value of $\Ge$ bounding from below the range where $\Pe_0$ exists for transverse modes $(\varphi = 0)$. From Fig.~\ref{fig6}, one may infer that the transverse modes are the most unstable among the modes with infinite wavelength if $5.99306 < \Ge \le 3\sqrt{5}$. A different scenario exists when  $0 < \Ge < 5.99306$ as the longitudinal modes turn out to be the most unstable.  

\section{Discussion of the results}
The neutral stability condition for the normal modes with infinite wavelength could be found through an analytical solution by employing a power series expansion with respect to the wavenumber $k$. Extending the scope to perturbation modes with a finite wavelength implies a numerical solution of \equas{18}.

\subsection{Numerical method}
The analysis carried out in Section~\ref{creeflo} highlighted that \equas{18} yields an eigenvalue problem. With either $A=A_{-}$ or $A=A_{+}$, one may consider $(k, \varphi, \Ge)$ as input parameters and obtain $(\Pe, \omega)$ as the eigenvalues. The strategy is that defined by the shooting method \cite{straughan2013energy, barletta2019routes}. The first step is setting up a numerical solver for the boundary value problem
\gat{
W'''' - 2\,k^2\, W'' + k^4\, W - k^2\, \Theta = 0 ,\label{30a}\\
\Theta'' - \qty[k^2 - i \omega + i k\, \Pe\, F(z) \cos\varphi ]\, \Theta + \frac{2 i\, \Ge\, \Pe\, F'(z) \cos\varphi}{k}\, \qty(W'' + k^2\, W) \nonumber\\
\hspace{2cm}
-\, \frac{i\, \Pe\, A \cos\varphi}{k}\, W' - \Pe^2\, G'(z)\, W = 0  , \label{30b}\\
W(0) = 0 \qc W'(0) = 0 \qc \Theta(0) = 1 \qc \Theta'(0) = 0 , \label{30c}\\
W(1) = 0 \qc W'(1) = 0 . \label{30d}
}{30}
The difference between \equas{18} and \equas{30} is that, in \equas{30}, we set $\lambda = - i \omega$, we do not impose the homogeneous boundary condition $\Theta'(1) = 0$, but we enforce the inhomogeneous boundary condition $\Theta(0) = 1$ instead, which is not present in \equas{18}. The latter inhomogeneous condition is legitimate as \equas{18} are scale invariant: if $(W, \Theta)$ is a solution, also $(C W, C \Theta)$ is a solution, for every complex constant $C$. Thus, fixing $\Theta(0) = 1$ means picking up a a single non--trivial solution $(W, \Theta)$ among an equivalence class of possible solutions. The constant $1$ can be replaced by any other complex number without affecting the solution of the eigenvalue problem (\ref{18}). This argument is just the same as that reported in Section~\ref{infwape} with reference to the eigenvalue problem (\ref{23}).

\begin{table}[t]
   \centering
   \begin{tabular}{c|l l | l l}
$\Ge$ & $\Pe_0\ (\varphi=\pi/2)$ & $\Pe\ (k=0.1, \varphi=\pi/2)$ & $\Pe_0\ (\varphi=0)$ & $\Pe\ (k=0.1, \varphi=0)$\\ 
   \hline\hline
   0.1 & 183.638 & 183.670 & --- & --- \\
   0.2 & 130.083 & 130.106 & --- & --- \\
   0.5 & \pg82.6889 & \pg82.7033 & --- & --- \\
   0.8 & \pg65.6755 & \pg65.6869 & --- & --- \\
   1 & \pg58.9108 & \pg58.9211 & --- & --- \\
   2 & \pg42.1479 & \pg42.1552 & --- & --- \\
   4 & \pg30.1012 & \pg30.1061 & --- & --- \\
   6 & \pg24.1831 & \pg24.1861 & 24.0109 & 24.0554 \\
   $3\sqrt{5}$ & \pg22.3114 & \pg22.3123 & 11.7021 & 11.7142 \\
\hline
   \end{tabular}
   \caption{\label{tab1}Comparison for the $A_{-}$ branch between the values of $\Pe_0$, obtained by the analytical expression (\ref{29}), and the neutral stability value of $\Pe$ with $k=0.1$ computed numerically by the shooting method} 
\end{table}

Equations~(\ref{30}) can be solved, with $A = A_{-}$ or $A = A_{+}$, for every fixed set of parameters $(k, \varphi, \Ge, \Pe, \omega)$. The numerical solution is accomplished by the software tool {\sl Mathematica} (\copyright{ Wolfram Research, Inc.})~via the built-in function {\tt NDSolve}.  Eventually, the condition  $\Theta'(1) = 0$ excluded from \equas{30}, but defining the eigenvalue problem (\ref{18}), is employed to determine the eigenvalues $(\Pe, \omega)$ for every input data $(k, \varphi, \Ge)$. In fact, such a homogeneous condition yields two constraints, $\Re[\Theta'(1)] = 0$ and $\Im[\Theta'(1)] = 0$, with $\Re$ and $\Im$ the real and imaginary parts. Thus, this condition leads to the determination of the two real parameters $(\Pe, \omega)$. The solution of the target constraints, $\Re[\Theta'(1)] = 0$ and $\Im[\Theta'(1)] = 0$, is accomplished by using the function {\tt FindRoot} of {\sl Mathematica}. By employing the analytical solution found in Section~\ref{infwape}, one can initialise the root finding algorithm for the evaluation of $(\Pe, \omega)$ by starting with $k \to 0$, for all cases where $\Pe_0$ is defined, and gradually increasing $k$. Equations~(\ref{24}) and (\ref{25}) imply that, in the limit $k \to 0$, $\omega$ is zero.

By exploring the domain of all possible input parameters $(k, \varphi, \Ge)$, one may carry out the linear stability analysis. A convenient representation of the results is displayed in the two-dimensional space $(k, \Pe)$, by drawing the neutral stability curve relative to a given pair $(\varphi, \Ge)$. Graphically, such a curve yields the condition of linear instability as that where $\Pe$ exceeds its minimum evaluated along the neutral stability curve. This condition of minimum $\Pe$ defines the critical values, $(k_c, \Pe_c, \omega_c)$, for the onset of the instability \cite{drazin2004hydrodynamic, straughan2013energy, kundu2016fluid, barletta2019routes}. 

A validation of the numerical solver is reported in Table~\ref{tab1} where the data for $\Pe$ obtained from the analytical expression (\ref{29}) are compared with those computed numerically by the shooting method for $k=0.1$. An excellent agreement is found if one considers that the values of $\Pe_0$ and the neutral stability values of $\Pe$ evaluated numerically are relative to slightly different wavenumbers $(k=0$ and $k=0.1)$. The data for transverse modes with $\Ge \le 4$ are not reported in Table~\ref{tab1} as $\Pe_0$ is undefined when $\Ge < 5.47464$ as specified in Section~\ref{infwape}.

\begin{figure}[t]
\centering
\includegraphics[width=0.41\textwidth]{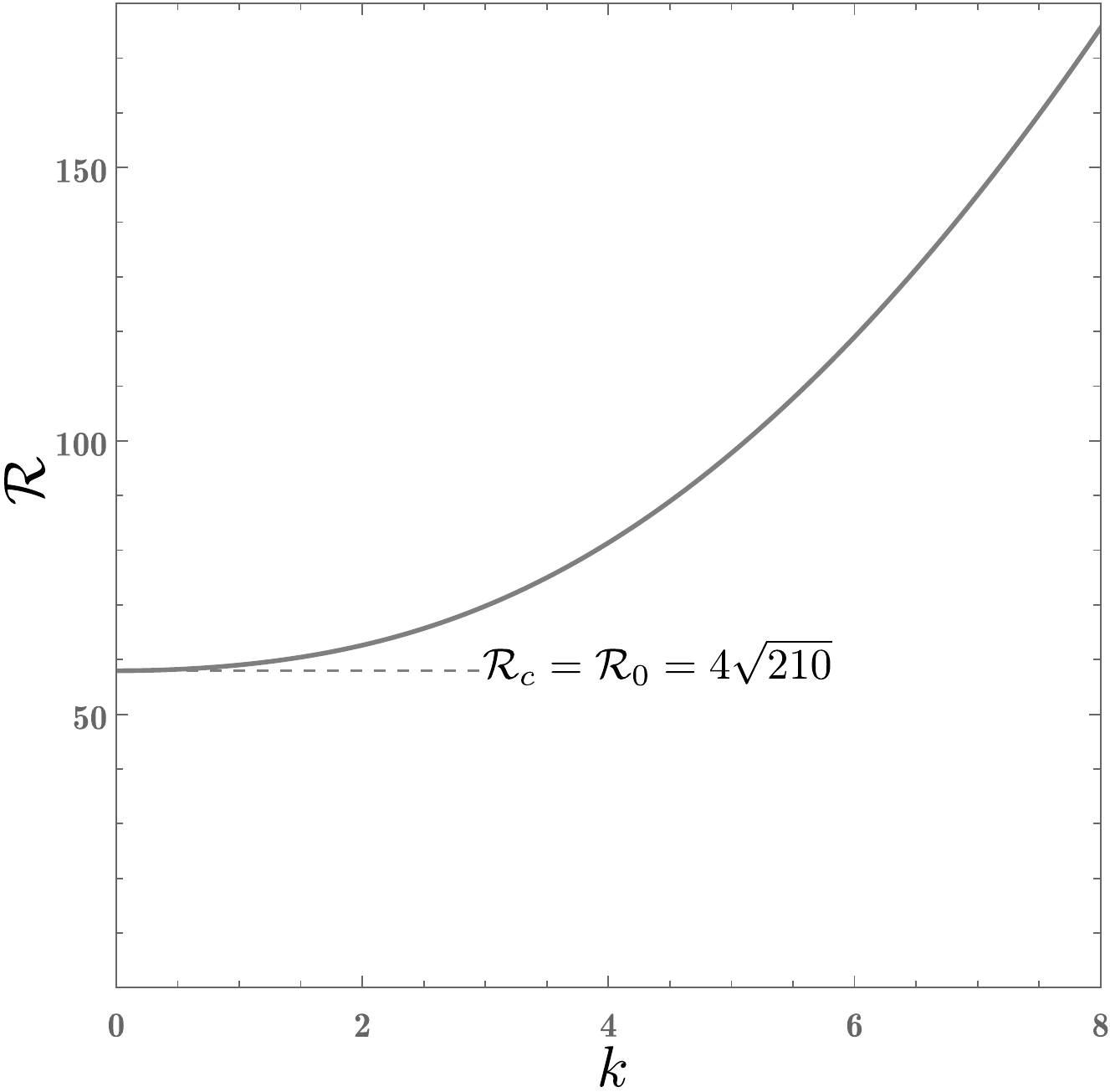}
\caption{\label{fig7}Neutral stability curve in the $(k, \mathcal{R})$ plane for longitudinal modes with $A = A_{-}$ and $\Ge \to 0$}
\end{figure}

\subsection{The regime of small Gebhart number}\label{smgenu}
One can investigate the asymptotic solution of the eigenvalue problem (\ref{18}) in the limit of small values of $\Ge$. By setting $A = A_{-}$, one can detect the behaviour at the lowest order in $\Ge$ from \equa{12}. Hence, for very small Gebhart numbers, \equas{18} can be approximated as
\gat{
W'''' - 2\,k^2\, W'' + k^4\, W - k^2\, \Theta = 0 ,\label{31a}\\
\Theta'' - \qty[k^2 - i \omega + i k\, \Pe\, z \cos\varphi ]\, \Theta + \frac{2 i\, \Ge\, \Pe \cos\varphi}{k}\, \qty(W'' + k^2\, W) \nonumber\\
\hspace{2cm}
-\, \frac{2 i\, \Ge\, \Pe \cos\varphi}{k}\, W' + \Ge\, \Pe^2\, z \qty(1 - z)\, W = 0  , \label{31b}\\
W = 0 \qc W' = 0 \qc \Theta' = 0 \qfor z = 0, 1. \label{31c}
}{31}
The first consideration is that the limit $\Ge \to 0$ can be taken by taking, contextually, also the limit $\Pe \to \infty$ (see the discussion of this point in \citet{doi0144878}). This double limit is well-defined provided that one considers 
\eqn{
\mathcal{R} = \Pe \sqrt{\Ge} \sim \order{1} .
}{32}
Thus, \equas{31} can be rewritten as
\gat{
W'''' - 2\,k^2\, W'' + k^4\, W - k^2\, \Theta = 0 ,\label{33a}\\
\Theta'' - \qty[k^2 - i \omega + i k\, \frac{\mathcal{R}}{\sqrt{\Ge}}\, z \cos\varphi ]\, \Theta + \frac{2 i\, \sqrt{\Ge}\, \mathcal{R} \cos\varphi}{k}\, \qty(W'' + k^2\, W) \nonumber\\
\hspace{2cm}
-\, \frac{2 i\, \sqrt{\Ge}\, \mathcal{R} \cos\varphi}{k}\, W' + \mathcal{R}^2\, z \qty(1 - z)\, W = 0  , \label{33b}\\
W = 0 \qc W' = 0 \qc \Theta' = 0 \qfor z = 0, 1. \label{33c}
}{33}
There are two possible outcomes for the limit $\Ge \to 0$ which depend on the inclination angle $\varphi$. If $\cos\varphi \ne 0$ or, equivalently, if $\varphi \ne \pi/2$,  the limit $\Ge \to 0$ with $\mathcal{R} \sim \order{1}$ yields a dominant term in \equa{33b}, of order $\Ge^{-1/2}$, so that this equation can be satisfied only with $\Theta = 0$. By substituting $\Theta = 0$ in \equa{33a}, the unique solution of \equasa{33a}{33c} is $W = 0$. In other words, there are no perturbation modes leading to a neutral stability condition with $\varphi \ne \pi/2$.
On the other hand, if $\varphi = \pi/2$, there exists a limiting formulation of \equas{33} for $\Ge \to 0$,
\gat{
W'''' - 2\,k^2\, W'' + k^4\, W - k^2\, \Theta = 0 ,\label{34a}\\
\Theta'' - \qty(k^2 - i \omega )\, \Theta + \mathcal{R}^2\, z \qty(1 - z)\, W = 0  , \label{34b}\\
W = 0 \qc W' = 0 \qc \Theta' = 0 \qfor z = 0, 1,\label{34c}
}{34}
which admits non--trivial solutions.

\begin{figure}[t]
\centering
\includegraphics[width=0.41\textwidth]{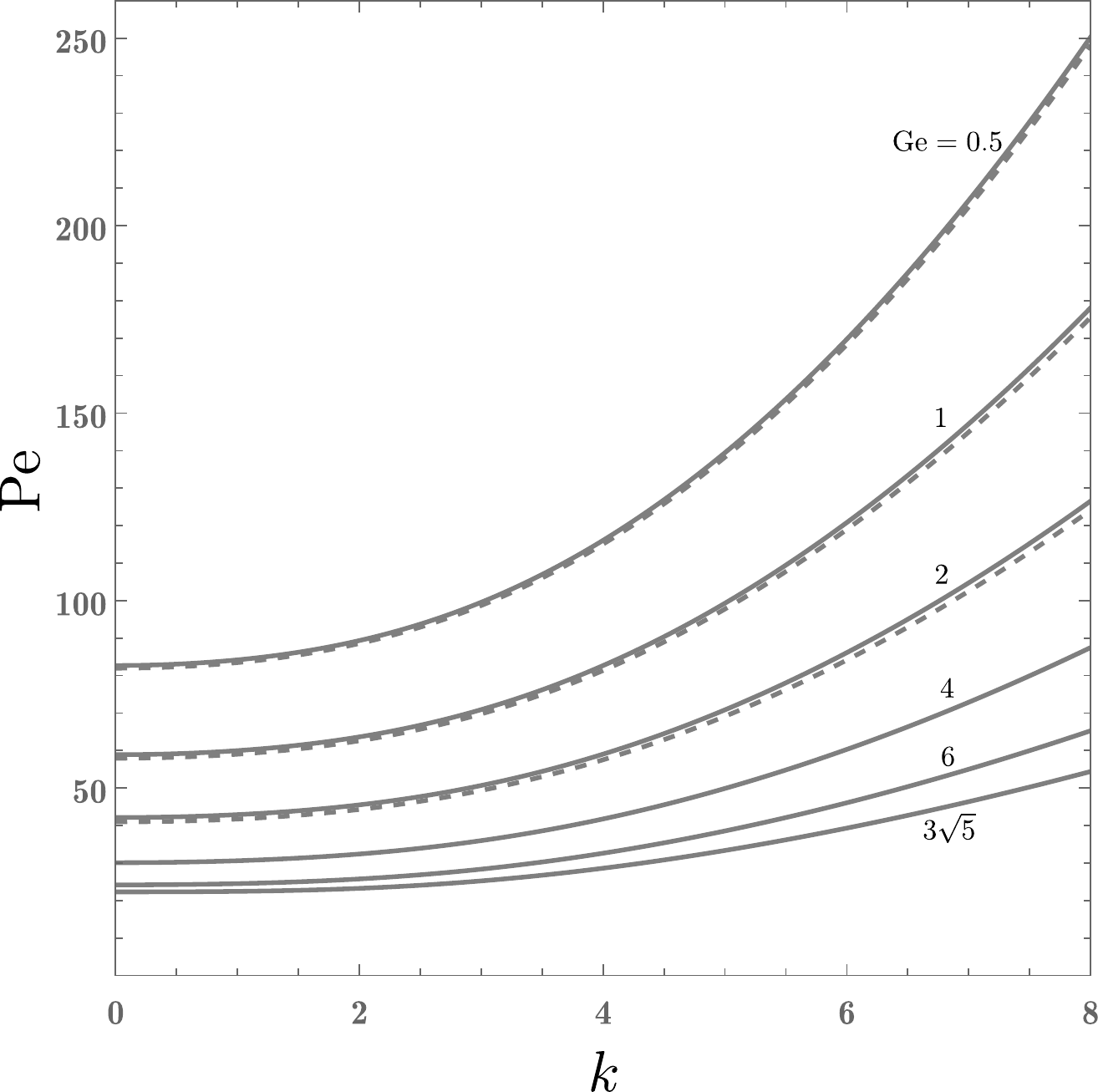}
\caption{\label{fig8}Neutral stability curves (solid lines) in the $(k, \Pe)$ plane for longitudinal modes with $A = A_{-}$ and $\Ge$ ranging from $0.5$ to its maximum value $3\sqrt{5}$. The dashed lines display, for $\Ge = 0.5, 1$ and $2$, the neutral stability data evaluated with the asymptotic solution for $\Ge \ll 1$}
\end{figure}

The eigenvalue problem (\ref{34}) may be solved numerically to determine the neutral stability curve for the longitudinal modes $(\varphi = \pi/2)$ in the $(k, \mathcal{R})$ plane. Such a curve is depicted in Fig.~\ref{fig7}. The first comment is that the neutral stability curve represents $\mathcal{R}$ as a monotonic increasing function of $k$, so that the critical value of $\mathcal{R}$ is 
\eqn{
\mathcal{R}_c = \mathcal{R}_0 = \lim_{\Ge \to 0} \Pe_0\, \sqrt{\Ge} = 4 \, \sqrt{210} \approx 57.9655 ,
}{35}
where the limit in \equa{35} has been evaluated by using \equa{29}. The role of \equa{35} is compelling as it provides an analytical expression that can be used to approximate the critical value of $\Pe$ for the most unstable modes, {\em viz.} the longitudinal modes, at small values of the Gebhart number,
\eqn{
\Pe_c = \Pe_0 \approx 4 \, \sqrt{\frac{210}{\Ge}} .
}{36}
Another important feature of the numerical solution of \equas{34} is that, along the neutral stability curve,
$\omega=0$. In fact, this result is a characteristic trait of longitudinal modes at both small and large Gebhart numbers.

\subsection{The most unstable perturbation modes}
The determination of the most unstable modes or, equivalently, the determination of the angle $\varphi$ that yields the lowest value of $\Pe_c$ for a given Gebhart number is a primary step in the stability analysis. In Section~\ref{smgenu}, we have concluded that, for the asymptotic condition $\Ge \ll 1$, the longitudinal modes lead the transition to linear instability with $\Pe_c$ evaluated analytically through \equa{36}. The oblique or transverse modes, having $\varphi \ne \pi/2$, do not yield any asymptotic condition of neutral stability in this asymptotic case. In practice, such a behaviour means that $\Pe_c$ for oblique or transverse modes tends to infinity much faster than for the longitudinal modes when $\Ge \to 0$. Thus, we figure out a scenario where, for small Gebhart numbers, the selected modes at the onset of the linear instability are longitudinal. In particular, the transition is activated by longitudinal modes with infinite wavelength, {\em i.e.} $\Pe_c = \Pe_0$. 

\begin{figure}[p]
\centering
\includegraphics[width=0.82\textwidth]{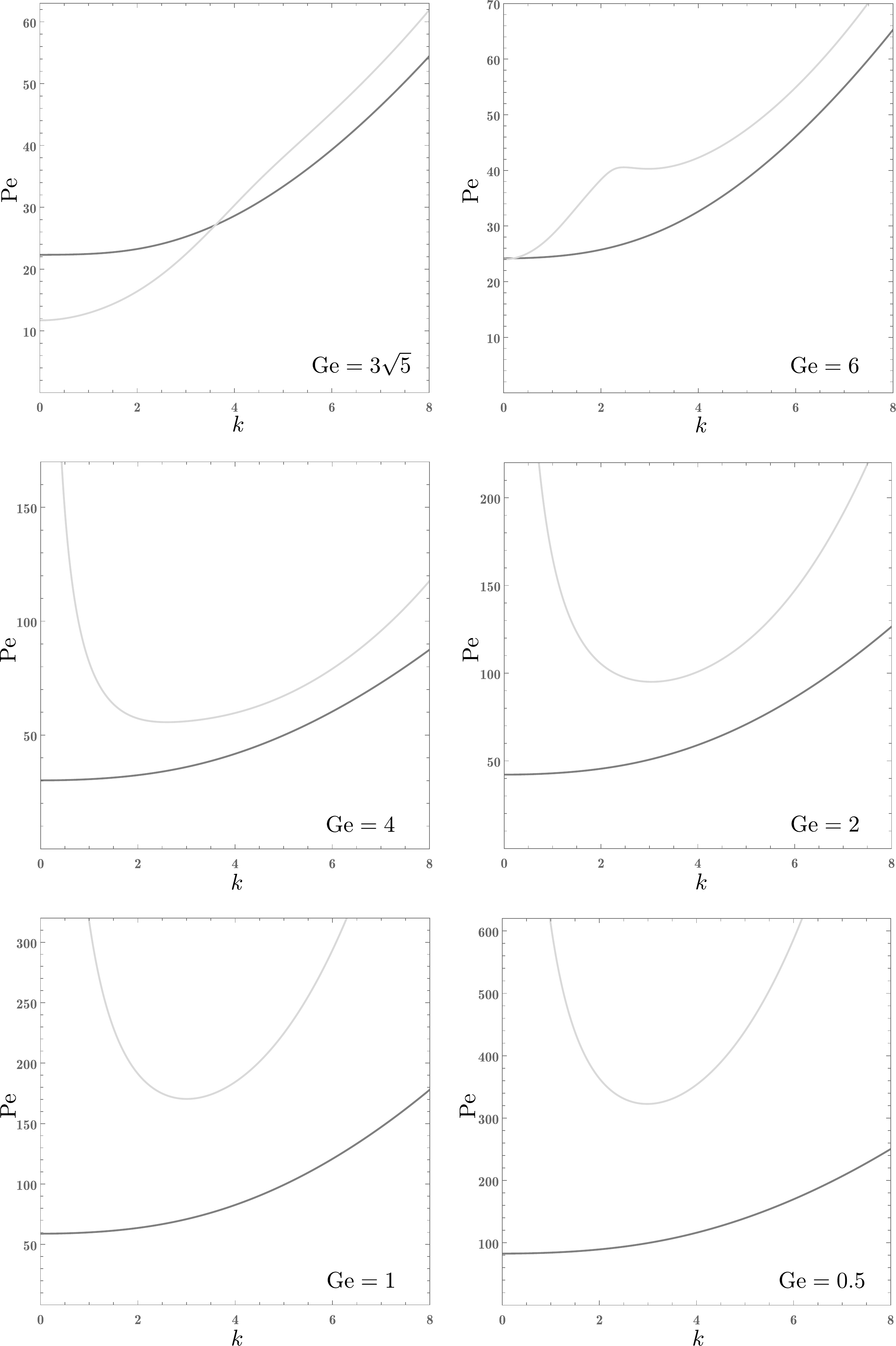}
\caption{\label{fig9}Neutral stability curves in the $(k, \Pe)$ plane with $A = A_{-}$ for longitudinal modes (dark grey lines) and transverse modes (light grey lines)}
\end{figure}

As illustrated in Fig.~\ref{fig8}, the critical value of $\Pe$ for longitudinal modes, relative to $A = A_{-}$, coincides with $\Pe_0$ whatever is the prescribed Gebhart number. In fact, all the neutral stability curves show $\Pe$ as a monotonic increasing function of $k$. By inspecting Fig.~\ref{fig8}, the neutral stability data reported in Fig.~\ref{fig7} for the limiting case $\Ge \ll 1$ turn out to determine, through the scaling $\Pe = \mathcal{R}/\sqrt{\Ge}$, the neutral stability condition with a rough accuracy for $\Ge = 2$ (the maximum relative discrepancy is $2.8\%$), a fair accuracy for $\Ge = 1$ (the maximum relative discrepancy is $1.6\%$) and an even better accuracy for $\Ge = 0.5$ (the maximum relative discrepancy is $0.86\%$). Hence, for practical purposes, the asymptotic solution with $\Ge \ll 1$ can be safely employed for all cases with $\Ge \le 0.5$.

\begin{figure}[p]
\centering
\includegraphics[width=0.82\textwidth]{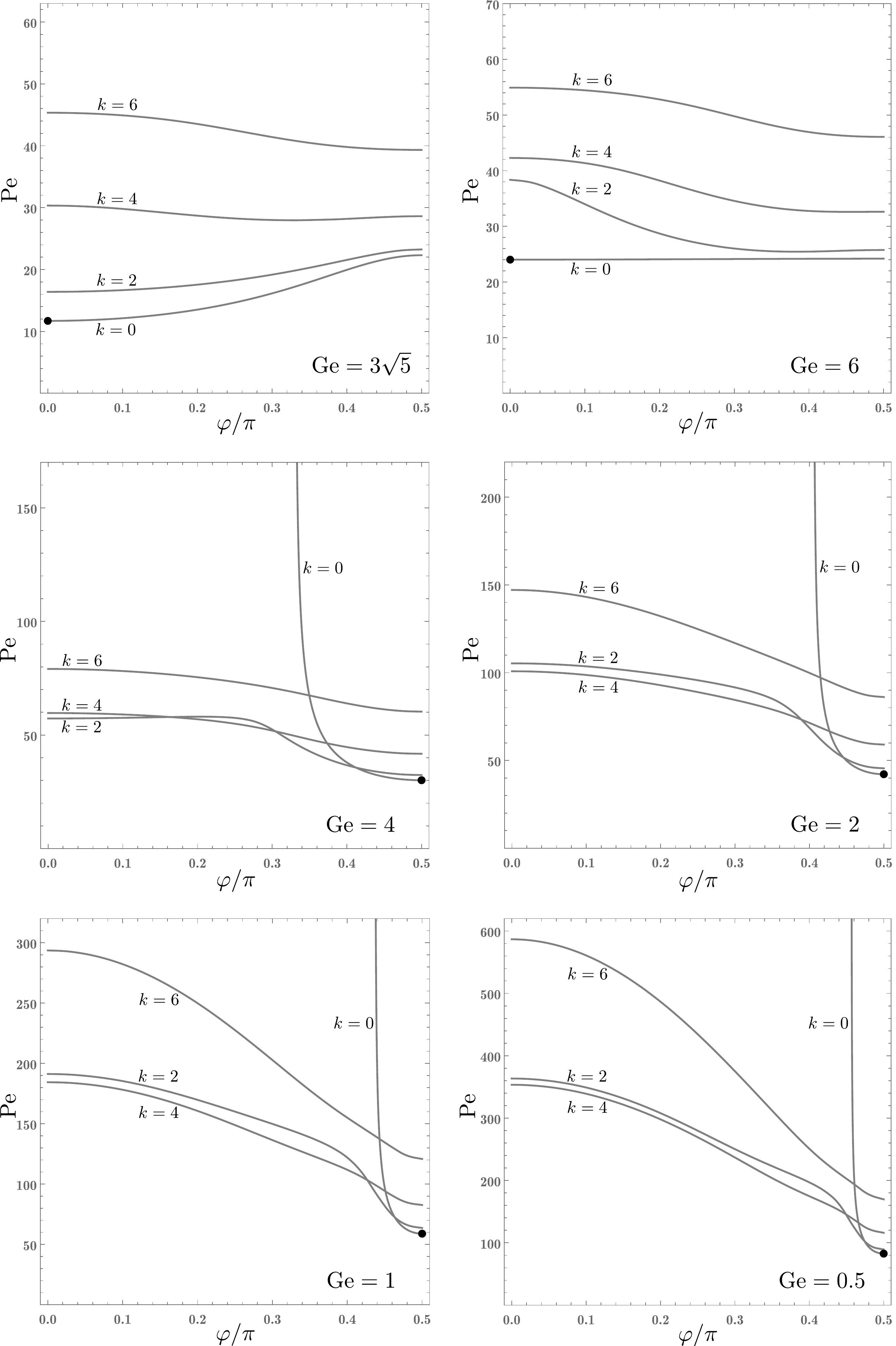}
\caption{\label{fig10}Neutral stability values of $\Pe$ versus $\varphi$ with $A = A_{-}$ for oblique modes at different wavenumbers. The black dots denote the minimum values of $\Pe$ for the transition to linear instability}
\end{figure}

As already highlighted in Section~\ref{infwape}, for infinite wavelength $(k \to 0)$, the longitudinal modes are not the most unstable when extremely large Gebhart numbers are considered and, in particular, when $\Ge > 5.99306$. Figure~\ref{fig9} provides a comparison of longitudinal and transverse modes beyond the quite specific condition $k \to 0$. This figure suggests that the behaviour conceived for the modes with infinite wavelength indeed holds in general. Indeed, Fig.~\ref{fig9} shows that the lowest minimum of the neutral stability curves for either longitudinal or transverse modes, for a given Gebhart number, is always at $k \to 0$. This means that the leading critical condition for the onset of the linear instability is that already discussed in Section~\ref{infwape}. Linear instability is triggered by transverse modes if $5.99306 < \Ge \le 3\sqrt{5}$, while it is started by longitudinal modes if $0 < \Ge < 5.99306$. Exploring Gebhart numbers below $\Ge = 0.5$ means just widening the gap between the neutral stability curve for longitudinal modes and that for transverse modes. 
Thus, one recovers the expected trend where the neutral stability threshold for transverse modes tends to an infinite $\Pe$ when $\Ge \to 0$ more rapidly than $1/\sqrt{\Ge}$, as anticipated with the analysis carried out in Section~\ref{smgenu}.

The intermediate conditions where $0 < \varphi < \pi/2$  can be inspected in an efficient way by tracking the change of $\Pe$ versus $\varphi$ with a given Gebhart number and wavenumber. Figure~\ref{fig10} displays a major result as it suggests that the transition to linear instability is always driven by the $k=0$ modes: either transverse at the extremely large Gebhart numbers close to the maximum, $\Ge = 3 \sqrt{5}$, or longitudinal when the Gebhart number is smaller than the threshold detected in Section~\ref{infwape}, namely for $\Ge < 5.99306$. In fact, we recall that $\Ge = 5.99306$ yields the special case where $\Pe_0$ is independent of $\varphi$, so that the onset of instability is triggered by any infinite wavelength modes whatever is their orientation in the horizontal plane. Then, we can take it as a general result that the critical P\'eclet number, $\Pe_c$,  for the initiation of the instability does always coincide with $\Pe_0$ for either $\varphi=0$ or $\varphi=\pi/2$ depending on the Gebhart number. Another feature displayed by Fig.~\ref{fig10} is that the neutral stability value of $\Pe$ may depend non-monotonically on $\varphi$ for a few cases as, for instance $k=2$ with either $\Ge=4$ or $6$.

\begin{figure}[t]
\centering
\includegraphics[width=0.82\textwidth]{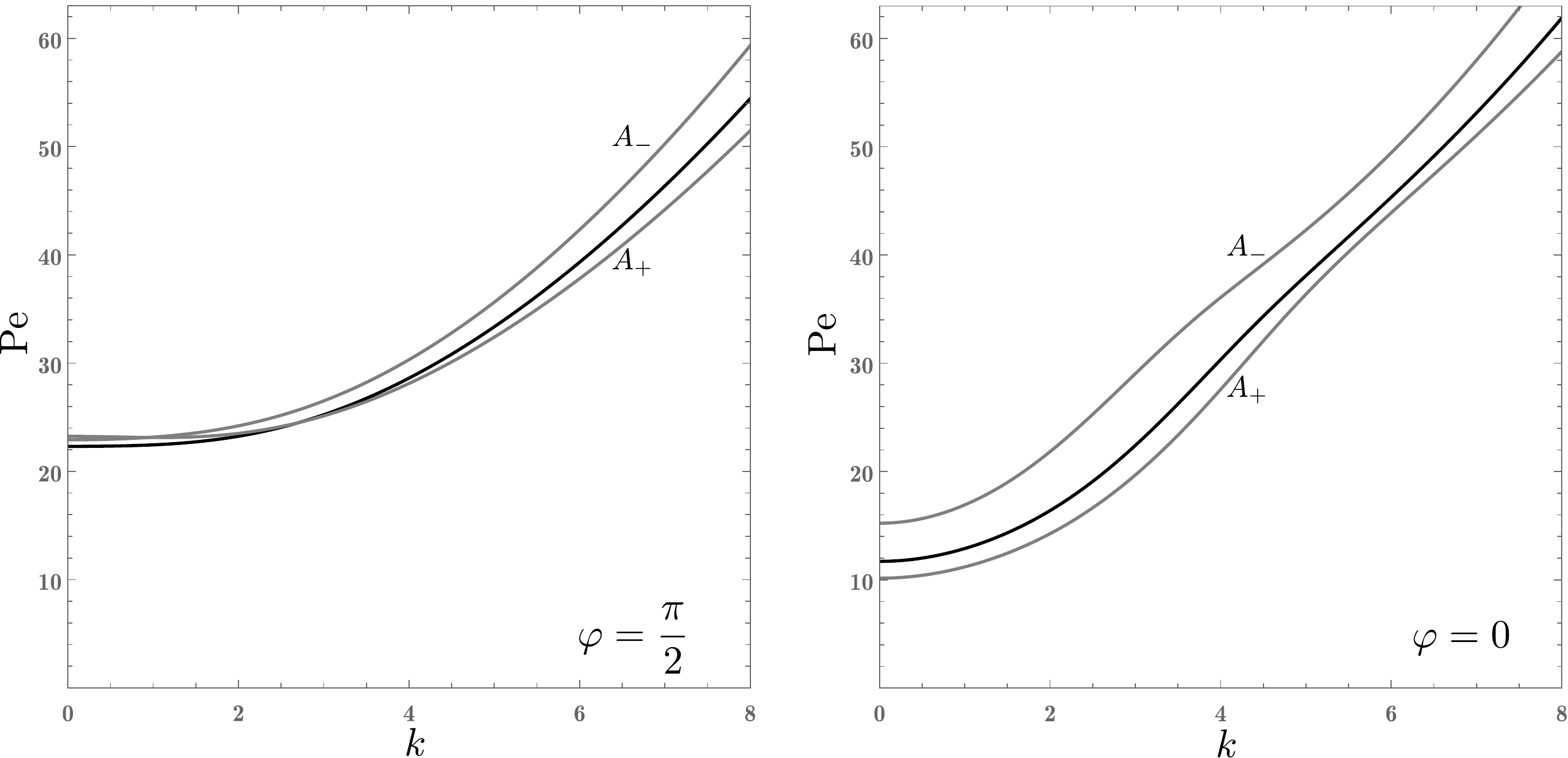}
\caption{\label{fig11}Neutral stability curves in the $(k, \Pe)$ plane for $\Ge=6.5$ (gray lines) and $\Ge = 3 \sqrt{5}$ (black line) relative to longitudinal modes $(\varphi=\pi/2)$ and transverse modes $(\varphi=0)$ with either $A = A_{-}$ or $A = A_{+}$}
\end{figure}

\subsection{Close to the maximum Gebhart number}
We have stressed that the maximum possible Gebhart number, $3\sqrt{5}$, is an extremely large value for most common applications, possibly also at large spatial scales such as those pertaining to geophysical systems. This said, it may be a matter of completeness in the stability analysis considering also the comparison between cases on the $A_{-}$ branch and the $A_{+}$ branch envisaged in Fig.~\ref{fig4}. An illustration of the transition to linear instability close to $\Ge = 3\sqrt{5}$ is displayed in Fig.~\ref{fig11}. Besides the expected similarity of the neutral stability curves for $A = A_{-}$ and $A = A_{+}$, an important result is that the transition to instability is driven by transverse modes with an infinite wavelength. Furthermore, the branch $A = A_{+}$ turns out to be more unstable than the branch $A = A_{-}$. Longitudinal modes are not involved in the initiation of the instability and they show up an uneven behaviour on comparing the two branches $A = A_{-}$ and $A = A_{+}$. In fact, the branch $A = A_{+}$ of longitudinal modes shows up a minimum condition for the P\'eclet number $\Pe_c = 23.1290$ with $k_c=1.07691$. Among the cases reported in Fig.~\ref{fig11}, the latter is the only one where the neutral stability curve involves a P\'eclet number not monotonically increasing with $k$.

\begin{table}[t]
   \centering
   \begin{tabular}{c| c}
   \hline\hline
kinematic viscosity, $\nu\ [{\rm m^2/s}]$  & $5.50 \times 10^{-4}$	  \\
thermal diffusivity, $\alpha\  [{\rm m^2/s}]$ & $8.53 \times 10^{-8}$  \\
specific heat, $c\ [{\rm J/(kg\, K)}]$  & 1\,910	  \\
thermal expansion coefficient, $\beta\ [{\rm 1/K}]$  & $7 \times 10^{-4}$ \\
Prandtl number & 6\,450  \\
\hline
   \end{tabular}
   \caption{\label{tab2}Thermophysical properties of unused engine oil  at an average temperature of $300\,{\rm K}$ \cite{bejan2004convection}} 
\end{table}

\subsection{A thought experiment}
The analysis carried out so far is entirely based on dimensionless quantities, but it may be interesting to check for possible applications or even conceivable experimental validations of the results in some specific cases.

We assumed from the beginning of the stability analysis that the fluid is to be considered as extremely viscous and with a low thermal diffusivity so that its Prandtl number can reasonably be considered as very large, consistently with the creeping flow assumption. An example can be an unused engine oil whose properties at an average temperature of $300\,{\rm K}$ are reported in Table~\ref{tab2}. Let us assume that a Couette experimental setup is designed with a distance $H = 1\, {\rm cm}$ between the walls in relative motion. Then, by employing the data reported in Table~\ref{tab2}, the Gebhart number turns out to be extremely small,
\eqn{
\Ge = 3.60 \times 10^{-8} .
}{37}
With such a small Gebhart number, the basic velocity profile is practically indistinguishable from the isothermal linear profile for the Couette flow. Furthermore, we can consistently employ \equa{36}, deduced for $\Ge \ll 1$, in order to evaluate the critical P\'eclet number leading to the linear instability of the basic flow,
\eqn{
\Pe_c = \Pe_0 \approx 4 \, \sqrt{\frac{210}{\Ge}} = 306\,000.
}{38}
As the transition to instability predicted by our study occurs by longitudinal perturbation modes with infinite wavelength superposed to the basic flow, the qualitative overall flow pattern changes from the straight parallel streamlines of the basic flow to the mutually oblique straight streamlines of the perturbed flow. We might also recall that, as the transition to instability occurs by the longitudinal modes $(\varphi = \pi/2)$, the streamwise direction of the basic flow is the $y$ axis. 
%

Finally, by employing Table~\ref{tab2}, one can determine the critical Reynolds number for the linear instability of the Couette flow,
\eqn{
\Rey_c = \frac{\Pe_c}{\Pr} = 47.4 .
}{39}
Interestingly enough, this small critical Reynolds number might be compared with those obtained for the isothermal Couette flow on evaluating the threshold for the nonlinear hydrodynamic stability via the energy method \cite{falsaperla2019nonlinear, falsaperla2022energy, giacobbe2022monotonic}. 

Let us employ the definition of the dimensional scale for the temperature, \equa{4}, and the expression of the dimensionless temperature distribution in the basic state given by \equass{8}{10}. Then, one may estimate that, with $\Ge = 3.60 \times 10^{-8}$, $\Pe = \Pe_c= 306\,000$ and $A=A_{-}$, the temperature gap between the lower and the upper wall at a given streamwise cross-section, $y = {\rm constant}$, is $3.8\, {\rm K}$. Moreover, the streamwise basic temperature gradient, $1.5 \times 10^{-4}\, {\rm K/m}$, is definitely negligible. 
Over such a temperature range, the standard formulation of the problem based on the Oberbeck-Boussinesq approximation may be considered as consistent. 

\section{Conclusions}
The stationary parallel flows in a horizontal plane channel with adiabatic rigid walls have been studied. The viscous dissipation effect caused by the imposed relative velocity between the boundary walls has been taken into account in the local energy balance. The flow description adopted and, in particular, the temperature coupling in the local momentum balance has been modelled according to the Oberbeck-Boussinesq approximation.  As a result, the basic flows turned out to be non-isothermal displaying Couette-like velocity profiles. It has been shown that there exist dual flow branches corresponding to given values of the P\'eclet number, $\Pe$, and of the Gebhart number, $\Ge$. The dual flows coincide when $\Ge = 3\sqrt{5}$. No stationary parallel flows exist when $\Ge > 3\sqrt{5}$.
It has been pointed out that only one of these dual flow branches, denoted as the $A_{-}$ branch, is compatible with the Oberbeck-Boussinesq approximation for realistic, {\em i.e.} sufficiently small, Gebhart numbers.

A linear stability analysis focussed on the $A_{-}$ branch has been carried out in a creeping flow regime where the Prandtl number has been considered infinite. Arbitrarily oriented wavelike perturbations have been regarded, ranging from the longitudinal modes propagating in a direction perpendicular to the basic flow direction to the transverse modes, whose direction of propagation is parallel to the basic flow direction. All intermediate inclinations, namely the oblique modes, have been also considered. The main conclusions drawn from such an analysis can be outlined as follows:
\begin{itemize}
\item A numerical solution of the stability eigenvalue problem has been obtained, based on the shooting method. The objective has been the determination of the neutral stability curves in the $(k, \Pe)$ plane. Hence, it has been shown that, in every case, the smallest neutrally stable P\'eclet number, {\em i.e.} the critical value of $\Pe$, corresponds to the limit $k \to 0$ (infinite wavelength).
\item The initiation of the instability occurs when the P\'eclet number becomes larger than its critical value, which depends on the Gebhart number. The most unstable perturbation modes have an infinite wavelength. They are either longitudinal modes, for $\Ge < 5.99306$, or transverse modes, for $5.99306 < \Ge \le 3\sqrt{5}$. The case $\Ge = 5.99306$ is special as the transition to instability is independent of the mode orientation, {\em i.e.} longitudinal, oblique and transverse modes are equivalent.
\item An analytical solution of the stability eigenvalue problem has been obtained for the asymptotic case where the wavenumber, $k$, is vanishingly small and, as a consequence, the wavelength tends to infinity. Thus, in every case, the determination of the critical P\'eclet number is analytical.
\item The regime of small Gebhart numbers has been explored by an asymptotic solution. This regime is typical of flows on a laboratory scale, while cases where the Gebhart number is of the order of unity can only be pertinent for the study of geophysical or astrophysical systems \cite{kincaid1996role, miyagoshi2013vigor}. As for the analogous case of Poiseuille-like flows \cite{doi0144878}, the $\Ge \ll 1$ asymptotic solution reveals that the neutral stability P\'eclet number for longitudinal modes scales with $\Ge^{-1/2}$. As expected from the proof by \citet{romanov1973stability}, the basic flows turn out to be stable at any P\'eclet number when $\Ge \to 0$.
\item Based on the $\Ge \ll 1$ asymptotic solution, an experimental setup has been proposed for a possible validation of the results relative to flows having a vertical width compatible with the size of a laboratory equipment.
\end{itemize}
There are several opportunities for future developments of the results obtained in this paper.
The extension of the classical Oberbeck-Boussinesq approximation for convective flows to cases where the fluid viscosity undergoes a sensible temperature change may be important. Another possible improvement in the model adopted is relaxing the assumption of creeping flow for the perturbation dynamics or, equivalently, extending the study to cases with a finite Prandtl number. The nonlinearity of the transition to instability is another important issue. Its analysis could disclose the emergence of a possible subcritical instability and provide an effective comparison with the energy method results for the hydrodynamic stability threshold.

 
\subsection*{Acknowledgement}
The authors acknowledge financial support from Italian Ministry of Education, University and Research (MIUR) grant number PRIN 2017F7KZWS.

\bibliographystyle{elsart-num-names}
\bibliography{biblio}

\end{document}